\documentclass{aa}
\usepackage{graphicx}
\begin{document}
   \title{ Physical conditions in the dwarf local irregular galaxy IC 10. I- Diffuse Ionized Gas \fnmsep\thanks{Based on observations made at San Pedro M\'artir Observatory, 
Baja California, Mexico}}

   \author{A.M. Hidalgo-G\'amez 
          \inst{1},\inst{2}
          }

   \offprints{Hidalgo-G\'amez, A.M.}

\institute{Instituto de Astronom\'{\i}a, Universidad Nacional Aut\'onoma de M\'exico, Mexico City, Mexico\\
              \email{anamaria@astroscu.unam.mx}
\and 
Dept F\'{\i}sica, Escuela Superior de F\'{\i}sica y Matem\'aticas, IPN, Mexico City, Mexico
\\
          }

   \date{Received ; accepted }

 \abstract{
A detailed analysis on the physical conditions of the Interstellar Medium in the main body of the Local Group dwarf irregular galaxy IC 10 is carried 
out using long-slit spectroscopy. Maps of the excitation, the H$\alpha$ flux, [NII]/H$\alpha$ and 
[SII]/H$\alpha$ are presented. The Diffuse Ionized Gas inside the galaxy is studied. We found that the spectral characteristics are not similar to those for spiral galaxies: the values of the ratios [NII]/H$\alpha$ and [SII]/H$\alpha$ are not as large as in spiral galaxies but the excitation is much larger. These values, especially the large excitation, can be explained by proposing an extra source of ionization. Different sources are studied but any of the models at hand could be really fitted except a combination of photons leaking from the H\,{\sc ii} regions in addition to the ionization provided by the WR stars present in this galaxy. Shocks are not needed for explaining the line intensity ratios except in a small part in the west side of the galaxy. 

 }

\titlerunning{Physical conditions in IC 10.I-Diffuse Ionized Gas} 
  \authorrunning{Hidalgo-G\'amez}

   \maketitle
%
\section{Introduction}

IC 10 is one of the most fascinating galaxies in the Local Group.
It is located in a relatively isolated region at 660 kpc from the
Galaxy (Sakai et al. 1999). This galaxy has a number of characteristics which all together make it unique: there is a large H\,{\sc i} envelope with  a 
radius ten times larger than the optical (Shostak \& Skillman 1989), which is corrotating  (Wilcots \& Miller 1998). It has a large number of H\,{\sc ii} regions (Hodge \& Lee 1990, hereafter HL90). Finally, it has the largest density of WR stars of any dwarf irregular in the neighbourhood  (Crowther et al. 2003). So far, $20$ WR stars have been detected but more than $100$ are suspected (Massey \& Holmes 2002). The main problem is the large foreground reddening due to its location very close to the galactic plane ($l = -3.3$) and therefore little reliable information has been available until
very recently (e.g. Hunter 2001; Richer et al. 2001; Borissova et al. 2000). 

These characteristics, especially the surface density of WR stars, make IC 10 a Rossetta stone for many studies in astronomy. One is the influence of the WR stars on the interstellar medium (ISM). It is believed that this type of stars will have a strong influence on 
the enviroment due to their strong winds. These winds might create bubbles 
and pollute the medium with the products of the chemical evolution of the stars. A larger non-thermal cavity associated with a Supernova remmant has been 
already 
discovered (Shostak \& Skillman 1989). Moreover, it is believed that the 
Diffuse Ionized Gas (DIG) is ionized by photoionization and shocks. If the 
winds from the WR stars can create bubbles then the shock waves can ionize 
the medium and the effects will be visible in the main body of the galaxy, where the majority of WR stars has been detected. Finally, the release of these elements can change the chemical abundances of the ISM.  

In the present investigation we have focussed on the study of the Diffuse Ionized Gas inside IC 10 with the help of long-slit spectroscopy, while the influence on the ISM of the WR stars and the chemical abundances of the galaxy will appear in an upcoming paper (Hidalgo-G\'amez, AJ submitted; hereafter paper II) . 

The paper is structured as follows: Description of the data acquisition  is presented in Section 2. The physical properties of the galaxy (excitation, surface brightness in H$\alpha$, etc..) are described in Section 3, while Section 4 studies the characteristics of the DIG. The possible sources of ionization for this galaxy are presented in Section 5. Conclusions are outlined in Section 6.

\section{Observations and data reduction}

The observations were obtained on 9-12th October, 1999, (UT) with the
2.1-m telescope of the Observatorio Astron\'omico Nacional at San
Pedro M\'artir (OAN-SPM), Baja California, Mexico.  The Boller
\& Chivens spectrograph was used with a 400 l/mm grating blazed at
5660\AA. The detector was a Thomson CCD with 15\,$\mu$m
pixels in a $2048 \times 2048$ format. The slit width was 400\,$\mu$m, subtending
5${\arcsec}$ on the sky, and yielding a spectral
resolution of 14-16 \AA.  The full spectral range observed was
3500-7000\AA~ but due to the high interstellar extinction
suffered by IC 10, as well as the low efficiency of the
detector at the blue end, the effective spectral range was
4000-7000\AA.

\begin{figure}
\centering
\includegraphics[width=8cm]{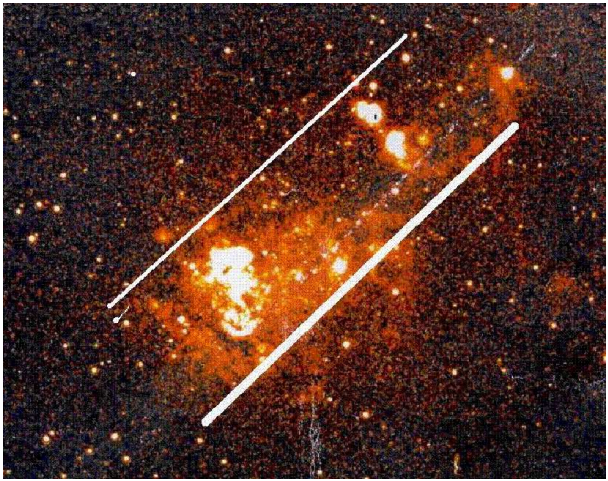}
\caption{H$\alpha$ image of the galaxy with the initial and the final positions of the slits marked. This image was taken by the author on September 1997 at the 2.5m NOT telescope with ALFOSC. The original field of view was 5'$\times$5' but the size presented here is smaller.}  
\label{ic10h}
\end{figure}

So, as to more efficiently map the star-forming area of IC 10, the slit
was oriented at a position angle of 139$^\circ$.  The different
positions within IC 10 were obtained by offsetting from the bright
star  GSC 3665-01255 $(\alpha = 0^h 20^m 20.9^s ,\,\delta =
59^\circ 17^\prime 36^{\prime\prime}$, J2000; HST Guide Star
catalogue, version 1.2).  In any given position, the telescope
guider was used to track a local guide star. Experience with this
technique and this telescope indicates that the offsets will be
uncertain to within approximately $\pm 2^{\prime\prime}$ in right
ascension. This problem was important for the first slit position of each night while the offsetting towards the east from it was perfect. Table 1 shows which slit positions were observed each night and the arcseconds, towards the east, the telescope was moved from the bright star of reference. A total of 15 spectra was obtained for the galaxy covering an 
area of $170$$\times$ $870$ pc, the major part of the inner galaxy (between the two white lines in Figure ~\ref{ic10h}). Three subexposures of $30$ minutes each were obtained for 
each slit position in order to check for cosmic ray events.

The data reduction was performed with the MIDAS software package. Bias and sky twilight 
flatfields were used for the calibration of the CCD response. One bias and 
flatfield were obtained for each night. He-Ar lamps were used for the wavelength calibration.  The spectra were corrected for atmosferic 
extinction using the San Pedro M\'artir tables (Schuster \& Parrao 2001). 
Several standard stars were 
observed each night in order to perform the flux calibration. The
accuracy of this calibration was from 1\% to 4\% for the four different 
nights, being the last night the poorest. Finally, the sky subtraction was 
a difficult task because of the large coverage of the 
emission lines through IC 10. For this reason, only a few rows at both ends of the spectra were useful and only the underlying sky structure was subtracted, but most of the stronger sky lines still remain in the spectra.  When a more severe subtraction was carried out, the signal from the lowest surface brightness part of the lines was removed. In the case of the third and fourth nights the situation became worse and only part of the underlying sky emission could be subtracted. Special care was taken in the analysis procedure for these nights. Therefore, the emission lines HeI $\lambda$5875 \AA~ and [OI]$\lambda$6300 \AA ~were not used in the present investigation.  Despite the large spectral range covered no atmospheric diffraction corrections were performed.

\begin{table}
\caption[]{ Log of observations. All the data were acquired with the 2.1m at OAN-San Pedro M\'artir Observatory with the use of the B\&Ch Spectrograph (See text for details). The first column shows the date of observation, and the second one the name of the slit position observed. The coordinates are listed in column 3. Here the first slit position is given by the exact coordinates but for the rest, only the number of arcseconds the telescope drifted from the initial position is given.   }
\vspace{0.05cm}
\begin{center}
\begin{tabular}{c c c}
\hline
{Night}   & {Slit} & {Position}    \\ 
\hline 
October 9th, 1999   & a & 0h 20m 21s   59$^o$ 17' 36"  \\
      & b & 5" towards the East     \\
      & c & 10" towards the East    \\
\hline
October 10th, 1999   & d & 16" towards the East    \\
      & e & 22" towards the East    \\
      & f & 28" towards the East    \\ 
      & g & 34" towards the East    \\
\hline
October 11th, 1999   & h & 40" towards the East    \\
      & i & 46" towards the East     \\
      & j & 52" towards the East     \\
      & k & 58" towards the East     \\
\hline
October 12th, 1999   & l & 64" towards the East     \\
      & m & 70" towards the East      \\
      & n & 76" towards the East      \\
      & o & 82" towards the East     \\       
\hline
\end{tabular}
\end{center}
\end{table}

A very important problem noticed during the reduction process was the existence of fringing. It is detected mainly beyond H$\alpha$ and is especially important at low surface brightness. In order to avoid this effect, all those pixels where the fringing could mask the real intensity of the lines were not considered in the study (see Section 4 for details).

The intensities of the lines were measured using the program PHYZ, developed at Uppsala Astronomical Observatory. 

Three different sources of uncertainties were considered: 
the uncertainties in the level of the spectral continuum with respect to the line,  $\sigma_c^2$, those introduced by the reduction procedure, $\sigma_r^2$, (especially flatfielding and flux calibrations) and uncertainties due to the extinction, $\sigma_e^2$.   The final uncertainty for 
each line was determined from $$\sigma = \sqrt {\sigma_c^2 + \sigma_r^2  + \sigma_e^2}$$  Due to the large number of spectra involved (more than $550$), an uncertainty was not associated with every line. Instead, the average of all the uncertainties for each line and for each slit position was determined and shown in Table 2. 

This set of data, fully reduced and calibrated, was used for the study proposed here.

\begin{table*}
\caption[]{ Total uncertainties in the most prominent emission lines in IC 10 for the 
DIG. Only one value was obtained for each line for each slit position. These values were 
determined using $\sigma = \sqrt {\sigma^2_{c} + \sigma^2_{r}  + 
\sigma^2_{e}}$ (see text for the meaning of the terms). The first column is the slit position labeled from the one positioned at the bright star and then moving it towards the east; the uncertainties in the 
emission lines are shown in the other columns. Those slits with uncertainties 
larger than $30$ \% in all the line ratios are considered to be of poor quality (slits a, l and n), while those with uncertainties smaller than $20$ \% are of good quality (slits named d, e, f, and i).  }
\vspace{0.05cm}
\begin{center}
\begin{tabular}{c c c c c  c}
\hline
{Slit }   & {H($\beta$)} & {[OIII]$\lambda$5007}  & {H($\alpha$)} & {[NII]$\lambda$6583} & {[SII]$\lambda$6716}  \\ 
\hline 
a & 45 \%   & 32 \%   & 33 \%   & 43 \%   & 33.4 \%  \\
b & 24 \%   & 29 \%   & 25 \%   & 27 \%   & 26 \%    \\
c & 22 \%   & 25.6 \% & 22 \%   & 22 \%   & 23 \%   \\
d & 19 \%   & 22 \%   & 20 \%   & 20 \%   & 18 \%    \\
e & 18 \%   & 18.6 \% & 18.3 \% & 16 \%   & 15 \%    \\
f & 17.5 \% & 18.5 \% & 21 \%   & 16 \%   & 17.5 \%  \\ 
g & 22 \%   & 22 \%   & 22 \%   & 23 \%   & 22 \%   \\
h & 22 \%   & 26 \%   & 19.5 \% & 20 \%   & 22 \%   \\
i & 18 \%   & 21 \%   & 17 \%   & 18 \%   & 17 \%   \\
j & 22 \%   & 26.7 \% & 20 \%   & 20.5 \% & 21 \%   \\
k & 20 \%   & 20 \%   & 25 \%   & 31 \%   & 20 \%  \\
l & 49.5 \% & 66.5 \% & 49 \%   & 49 \%   & 49 \%    \\
m & 27 \%   & 32 \%   & 25 \%   & 25.5 \% & 25.5 \%  \\
n & 31 \%   & 38.5 \% & 30 \%   & 30.5 \% & 32 \%    \\
o & 25 \%   & 25 \%   & 26 \%   & 34 \%   & 25 \%  \\       
\hline
\end{tabular}
\end{center}
\end{table*}

\section{Physical conditions in IC 10}

Before a more careful study of the properties of the DIG is considered, a global look at some of the characteristics of IC 10 will be presented.  In 
order to do it, all the 15 spectra for IC 10 were divided into 
nine-rows one-dimension spectra (hereafter 9r-spectra). This spatial 
resolution matches the slit width and corresponds to 19 parsec at the distance of IC 10. For each of the 9r-spectra the intensities of the Balmer 
lines H$\alpha$ and H$\beta$ as well as the 
forbidden lines [OIII]$\lambda$5007, [NII]$\lambda$6583 and [SII]$\lambda$6716+$\lambda$6731, were measured.  [OII]$\lambda$3727 was detected only at a very few locations with a very low S/N due to the small 
efficiency of the optics below 4000 \AA, so it will not be considered in this 
study. The lines [OI]$\lambda$6300 \AA~ and HeI $\lambda$5875 \AA~ were not used due to the poor sky subtraction. Another helium line, HeI $\lambda$6678 \AA, was also 
detected but due to the fringing problems that affected this line the most, its intensity might be largely masked.

The intensities for each one of these lines were normalized to H$\beta$ and corrected for extinction using the Balmer 
decrement with a value of $2.86$, corresponding to case B and an electronic temperature of $10,000$ K (Brocklehurst 1971). The extinction law used was the Whitford modified (Savage \& Mathis 1979). Non-absorption corrections were performed because they were important for only few locations they were finally rejected.

After these corrections, the ratios [OIII]/H$\beta$, [NII]/H$\alpha$ and [SII]/H$\alpha$ were obtained. Also, the surface brightness (SB) in the H$\alpha$ line was measured for the 9r-spectra. Four two-dimension maps, one for each of the ratios and the SB in H$\alpha$, were obtained from these data. In these maps all the 9r-spectra were considered, even those affected by fringing. In order to  discover differences, if any, throughout the galaxy, the maps were divided into the four cardinal points. The averaged values of the line ratios for the galaxy as a whole, the H\,{\sc ii} regions and the DIG locations (See next section for a definition) for all the galaxy and for the different cardinal points are presented in Table 3. Only fringing-free data were used in the determination of these values in the table.

\begin{figure}
\centering
\includegraphics[width=8cm]{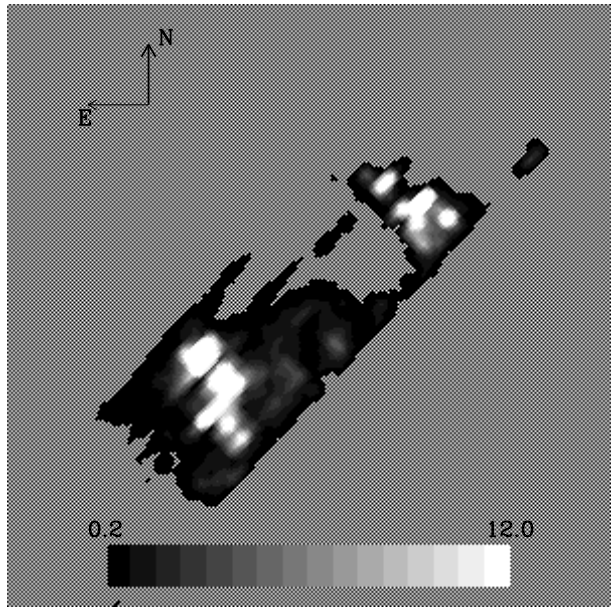}
\caption{This figure shows the surface brightness in the recombination line H${\alpha}$ for all the locations inside IC 10. 
North is up and east is to the left. 
This image is optimized in order to enhance the brightest parts of the galaxy. 
These correspond to surface brightness of $0.2$ to $12 \times 10^{-17}$ erg 
cm$^{-2}$s$^{-1}$ arcsec$^{-2}$ for the 9r-spectra. The
grey-scale bar is linear, in units of 10$^{-17}$ erg 
cm$^{-2}$s$^{-1}$ arcsec$^{-2}$. The background level is highlighted.
Although this map is created from spectra, the similarities with 
the H$\alpha$ image in Figure ~\ref{ic10h} are evident.}
\label{halfamap}
\end{figure}
  
Figure ~\ref{halfamap} shows the H${\alpha}$ flux map corrected for galactic extinction. The first thing to notice is that almost half of the galaxy has very low emission in H$\alpha$. This is clearly seen in the low average SB of the galaxy, which is only $1.81 \times 10^{-15}$ erg cm$^{-2}$ s$^{-1}$ arcsec$^{-2}$, as seen in Table 3. This agrees with the conclusions from HL90 that IC 10 is not suffering an intense burst of star formation extending to the whole galaxy. According to the most recent definition of starburst (Kennicutt 2004), the Star Formation Rate should be at least $10$ M$_{\odot}$ yr$^{-1}$ over a significant portion of the galaxy. For IC 10, values of only $1$ M$_{\odot}$ yr$^{-1}$ are obtained (Zucker 2005), that is a factor of ten smaller. There are other definitions of starburst galaxies. A galaxy can be classified as a real starburst if 
satisfies any of the following conditions: (i) $M_{*}/SFR \ll t_{Hubbl}$, 
(ii) its current Star Formation Rate is larger than five times the average SFR, (iii) the Star Formation is, at least, $100$ times  higher than in normal galaxies (Heckman 2005 and references therein). None of these definitions can qualify IC 10 as a bona fide starburst galaxy but only marginally, despite the blue color of this galaxy (Richer et al. 2001) and its WR population (e.g. Massey \& Holmes 2002).

According to results on Table 3, the western part of the galaxy has the total largest SB in H$\alpha$ while the northen region has the lowest. The south and the east, where the large and bright complex nr. $111$ is located, have intermediate fluxes between these two. As we divided our maps into 
the four cardinal points and this complex is very large, not all 
the spectra which belong to it were located at the same quadrant. 
Indeed, we noticed that the brightest region in this complex lays 
at the east, while the rest of the complex lay at the south of IC 10.  
However, when all the spectra corresponding to this region are 
considered together, this is the brightest region of the galaxy, 
as in HL90.

\begin{figure}
   \centering
   \includegraphics[width=8cm]{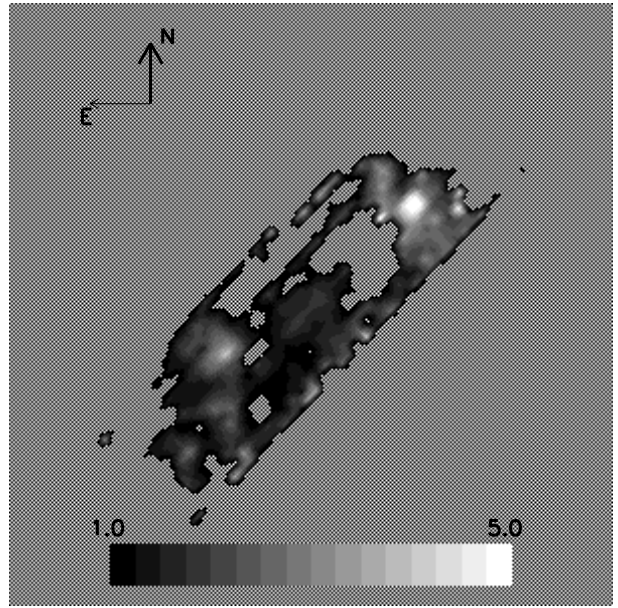}
     \caption{  This figure shows the excitation map defined as the value 
of [OIII]$\lambda$5007/H$\beta$. The  region of higher excitation 
corresponds to  H\,{\sc ii} region nr. 45. The orientation is the same 
as for the H$\alpha$ map.  The brightest parts of this figure correspond 
to excitation of $\approx 5$ while the data points with excitation lower 
than $1.0$ are those in black. The grey-scale is linear.}
        \label{oiiimap}
  \end{figure}

\begin{table*}
\caption[]{ Average values of the most prominent line ratios. The first 
column indicates the region considered: IC 10 means the average value for 
all the fringing-free data-points covering most of the optical body of the galaxy. H\,{\sc ii} and DIG are the same but considering only the values for 
the H\,{\sc ii} and DIG locations, respectively. In all the cases a subdivision into the four cardinal points is considered. The second column shows the 
SB in H$\alpha$ in units of $10^{-18}$ erg cm$^{-2}$ s$^{-1}$ arcsec$^{-2}$. 
The excitation is presented in column 5, while [NII]/H$\alpha$ 
and [SII]/H$\alpha$ 
are shown in columns 7 and 9. Finally, columns 3, 6, 8, and 10 list the number of 9r-spectra used for the determination of each value.}
\vspace{0.05cm}
\begin{center}
\begin{tabular}{c c c c c c c c c c c c}
\hline
{ }   & {SB(H($\alpha$))} & N & SB/N & {[OIII]$\lambda$5007/H$\beta$}   & N & {[NII]$\lambda$6583/H$\alpha$} & N &  {[SII]/H$\alpha$}  & N \\ 
\hline 
IC 10 & 1808.64 & 288 & 6.28 &1.86 & 285 & 0.16   & 139 &0.28& 228&\\
\hline
S & 502.88 & 112 & 4.49 & 1.61 & 109 & 0.17 & 61 & 0.36 & 94 \\
E & 553.04 & 62  & 8.92 & 3.87 & 62  & 0.11 & 33 & 0.28 & 50 \\
N & 109.12 &  44 & 2.48 & 1.87 & 44  & 0.09 & 13 & 0.17 & 32 \\
W & 655.92 & 72  & 9.11 & 2.24 & 70  & 0.11 & 32 & 0.20 & 52 \\
\hline
H\,{\sc ii} &1467. 44 & 68 & 21.58 & 2.05  & 69& 0.12  &67& 0.19 &61\\
\hline
S & 350   & 25 &  25   & 1.68   & 25& 0.18   &25 & 0.27  &25  \\
E & 495.9 & 19 & 26.10 & 1.72   & 20& 0.09   &20 & 0.14  &17\\
N & 61.2  &  3 & 20.40 & 2.54   & 3 & 0.06   &3  & 0.09  & 3\\
W & 560.7 & 21 & 26.7  & 2.74   & 21& 0.07   &19 & 0.12  &16  \\
\hline 
DIG & 352. 98 & 222 & 1.59 & 1.66  & 216&0.19  &72& 0.31 &167\\
\hline
S & 153.12  & 87  & 1.76 & 1.59  &84 & 0.17 &36& 0.39 &69  \\
E & 57.19   & 43  & 1.33 & 1.25  &42 & 0.13 &13& 0.36 &33 \\
N & 47.97   & 41  & 1.17 & 1.82  &41 & 0.10 &10& 0.18 &29 \\
W & 95.37   & 51  & 1.87 & 2.02  &49 & 0.16 &13& 0.24 &36\\
\hline
\end{tabular}
\end{center}
\end{table*}

A similar map is obtained for the excitation, defined as the [OIII]$\lambda$5007/H$\beta$ ratio (Fig. ~\ref{oiiimap}). From the H$\alpha$ map one might think that those high excitation regions are going to be very rare, and there will be very low values for the rest of the galaxy because the OB stars, which excite the gas, are not numerous. From Figure ~\ref{oiiimap} can be seen  that this is not the situation. The excitation is larger than expected. Actually, the total value for the galaxy is $1.86$ ($\sigma$ = $0.34$), only a factor of two smaller than the excitation in H\,{\sc ii} regions in this type of galaxies (Hidalgo-G\'amez \& Olofsson 2002). Moreover, it follows an uncommon behaviour. Locations with either very high ($>$ $2$) or very low ($<$ $0.5$) excitation are very rare, and most of the spectra have excitation between $1.8$ and $1$. This must indicate that the excitation inside IC 10 is not completely related with OB associations and that a part of it comes from another source. 

\begin{figure}
  \centering
   \includegraphics[width=8cm]{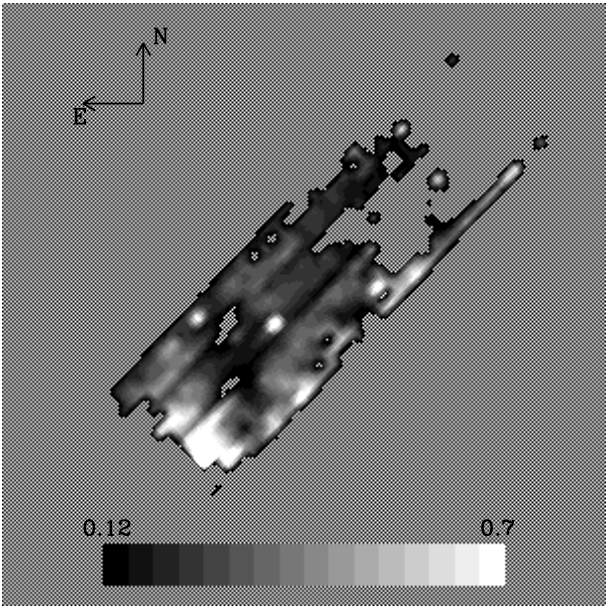}
\caption{ This figure shows the [SII]$\lambda\lambda$6717,6731/H$\alpha$ ratio. The orientation is as in Fig~\ref{halfamap}. It is interesting to notice the low values  in the vicinity of the H\,{\sc ii} regions. The limits in this figure ranges between $0.7$ (the brightest spots to the west) to $0.12$ (the darkest parts). The grey-scale is linear.} 
    \label{choques}
   \end{figure}

Figure ~\ref{choques} shows the map of the [SII]$\lambda$6717+6731/H$\alpha$ ratio. Though these lines are affected by fringing and some of the values are not 
correct, two important features are real. First, the low value of the 
averaged [SII]/H$\alpha$ ratio over the whole galaxy, which is $0.28$, as 
shows in Table 3. Second, and more important, are the very low values of this  ratio in the vicinity of the most prominent H\,{\sc ii} regions (See Fig. ~\ref{halfamap} and ~\ref{choques}). These results are real and not due to unrealistic values because of the fringing problem. It has been proposed that winds from OB associations can play an important role in the dynamics of the ISM, creating bubbles around these massive stars (e.g. Martin 1998). These bubbles will produce shock waves that could be detected from the [SII]/H$\alpha$, because this ratio is related to shocks (Dopita 1993). Figure  ~\ref{choques} indicates that these shocks are not detected in IC 10. Several explanations are available. It seems, from the results on the H$\alpha$ emission and the excitation, that OB associations in this galaxy are small. Therefore, they might not be able to create such bubbles. Another explanation is that they exist but cannot be detected mainly because sulphur is doubly ionized and, therefore, the ratio [SII]/H$\alpha$ is lower than expected. Another important piece of information about this ratio can be obtained from Table 3 and it is the existence of a gradient in this ratio, from south to north and from east to west. This result is also obtained with the fringing-free data. 

Finally, the map showing the [NII]$\lambda$6583/H$\alpha$ ratio is not presented here. The main reason is that this line is the most affected one by fringing, therefore the number of data-points which are realiable is very small and the results could not be real. However, there are a few remarkable results, confirmed by the data in Table 3. First, the noticeable large value of the [NII]/H$\alpha$ ratio for the whole galaxy, $0.16$, as well as in the H\,{\sc ii} regions ($0.12$). The former is a factor of $2.5$ larger than the typical values in some other irregular galaxies (Hidalgo-G\'amez \& Olofsson 2002). Moreover, it is quite homogeneous, at least at these large scales, except towards the south. 

An important caveat appears concerning the data. All these results could be an artifact because of the large uncertainties associated with the measurements due to the low S/N of the 9r-spectra. In order to check this, the set of data was sorted into three groups: those data-points 
corresponding to slits $d$, $e$, $f$ and $i$, that is those with the smaller uncertainties, slits $l$, $n$ and $a$  with the larger uncertainties, and the rest with medium uncertainties. The three ratios show similar values between the slits with large uncertainties ($2.06$, $0.4$ and $0.1$ for  [OIII]/H$\beta$, [SII]/H$\alpha$ and [NII]/H$\alpha$, respectively) and those with the smaller uncertainties ($1.45$, $0.32$ and $0.12$). Therefore, the low S/N locations do not have  a strong influence on the final results. In any case, all of the properties of these ratios, (e.g., high values of the excitation, low values of [SII]/H$\alpha$ and [NII]/H$\alpha$, etc) remain 
when only the data-points with the smaller uncertainties are considered. 

\section {Diffuse Ionized Gas in IC 10}

The Diffuse Ionized Gas (DIG) has been extensively studied in spiral 
galaxies, especially in the last 20 years.
Its main characteristics are a low excitation ([OIII]/H$\beta$ $<$ 0.2) and 
a 
low density (n$_e$ $\approx$ 10 cm$^{-3}$) (Mathis 1986). Also, very large values ($> 0.5$) of the [SII]/H$\alpha$ and [NII]/H$\alpha$ ratios (e. g. Rand 1998), and  of ~$5-10$ for the ratio [OII]/[OIII] (T\"ullman \& Dettmar 2000) are detected. 

Spiral galaxies are the main targets in DIG studies, mainly because the DIG is localized above the plane of the disk. The situation is more difficult for irregular galaxies where the disk is not clearly defined. For investigations such as the one presented here, where  a comparison between the properties of DIG and H\,{\sc ii} locations is carried out, a clear distinction between them is needed.

As said before, one of the main characteristic of the DIG is its low density. In the present investigation it is not possible to make use of the density as a discriminanting parameter because of the large uncertainties associated with the [SII] lines and the small differences in the [SII]$\lambda$6717/[SII]$\lambda$6731 ratio between H\,{\sc ii} regions and DIG.

The other parameter commonly used for a definition of DIG is the emission measure ($EM$) (Greenawalt et al. 1997) related with the surface brightness in H$\alpha$ by the expression $$EM = 7.22~ 10^{13} \times SB(H\alpha) \times T^{0.96}$$ where $T$ is the electronic 
temperature (typically $10,000$ K, e.g. Torres-Peimbert et al. 1974) and $SB(H\alpha)$ is the surface brightness in H$\alpha$ in erg cm$^{-2}$ s$^{-1}$ arcsec$^{-2}$. The values of $EM$ for the Galactic DIG is between $2$ and $80$ cm$^{-6}$ pc (Reynolds 1984) while values ranging between $10$ and $60$ are considered for  M31 (Greenawalt et al. 97). Hoopes \& Walterbos (2003) in their study of a few nearby spirals  considered DIG when $EM$ is lower than $100$ cm$^{-6}$ pc in the arms of spiral galaxies and below $50$ cm$^{-6}$ pc in the interarm regions.

\begin{figure}
\centering
\includegraphics[width=8cm]{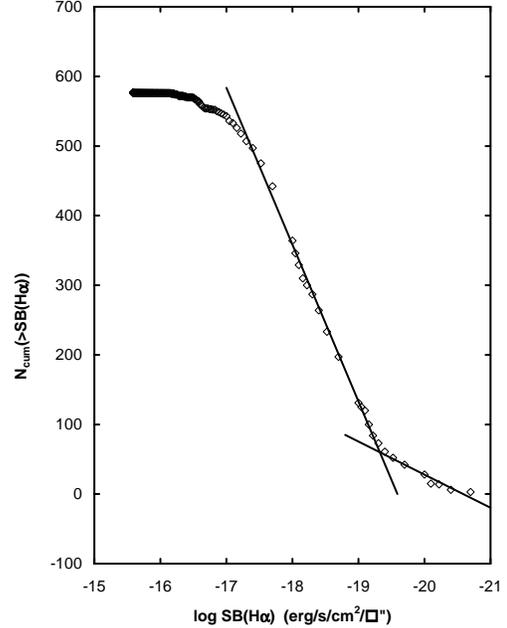}
\caption{This figure shows the cumulative distribution function of the SB in H$\alpha$ (galactic extinction corrected) for all the data points in IC 10. The lines are the best fitting to the data. Those data points with $\log(SB(H\alpha)) > -17.8$ are considered as H\,{\sc ii} regions while all those with $\log(SB(H\alpha)) < -19.3$ will be disregarded as noise level. } 
\label{histo}
 \end{figure}

In the present investigation another approach has been used. In order to have a definitive cut between H\,{\sc ii} and DIG, instead of choosing a cut value for the EM, the cumulative distribution function of the surface brightness for a total of $561$ data points in IC 10 is given in Figure  ~\ref{histo}. Two turn-off points, located at $\log (SB) ~\approx -17.4$ and at $\log (SB)~ \approx -19.3$, are clear. These values correspond to fluxes of $4~10^{-18}$ and $5 \times 10^{-20}$ erg cm$^{-2}$ s$^{-1}$ arcsec$^{-2}$, respectively, when corrected for galactic extinction using $E(B-V) = 0.77$ mag (Richer et al. 2001). 

Due to the problems with the fringing and the poor sky subtraction, it is likely that locations with very low intensities are going to be affected by them and their values might not be realiable. In order to avoid these poor quality and probably erroneous values we decided to increase the low limit up to $5 \times 10^{-19}$ erg cm$^{-2}$ s$^{-1}$ arcsec$^{-2}$. The number of data points is smaller but the results will be more trustful.  Therefore, in the following we will consider DIG locations inside IC 10 when the fluxes are between $5$ and $0.5 \times 10^{-18}$ erg cm$^{-2}$ s$^{-1}$. Locations with greater values corresponding to $EM$ larger than $2$ pc cm$^{-6}$ will be considered as H\,{\sc ii} locations. It is noticeable the large number of DIG locations (222) against only 68  H\,{\sc ii} locations. Finally, a  total of 273 data points were disregarded as noise level.

We can use Table 3 again in order to study the distribution of the ratios for the DIG and H\,{\sc ii} locations. The latter shows a similar distribution to that for the galaxy as a whole in the SB, with the largest values at the west and the lowest at the north. On the contrary, the southern DIG locations have the largest SB values while the northern ones have the lowest.

Another important feature is the large excitation ratio for DIG locations which is larger than for spiral galaxies (Greenawalt et al. 97). Actually, the values of the excitation for DIG are not very different from those for H\,{\sc ii} locations. Such large values and the small differences between DIG and H\,{\sc ii} cannot be explained by photoionization models, which always predict extremly low values for the [OIII]/H$\beta$ ratio (D\"omgonge \& Mathis 1994).  It is very interesting to notice that the largest excitation regions do not coincide with the largest SB ones, neither for H\,{\sc ii} nor the DIG locations. In both  cases the largest values come from the west and the lowest from the east. This could be, again, another indication that the excitation inside IC 10 is not completely related with the OB stars distribution. 

[SII]/H$\alpha$ is larger for DIG locations, in agreement with the photoionization  models. However, this ratio is very large in spiral galaxies: e.g., $0.7$ in the Milky Way (Reynolds 1984) and $0.5$ in M31 (Greenawalt et al. 1997). In IC 10, only $20$ out of the $156$ DIG locations available present [SII]/H$\alpha$ $>$ $0.5$. Though Figure ~\ref{choques} shows that the [SII]/H$\alpha$ is very low in the vicinity of the H\,{\sc ii} regions, Table 3 indicates that, on average, the south and the east have the largest values of this ratio. 

Finally, the [NII]/H$\alpha$ ratio gives very interesting results. Photoionization models (Mathis 1986; D\"omgonge \& Mathis 1994) predict a constancy of this ratio between H\,{\sc ii} regions and DIG. Such a constancy is not found for most of the galaxies, e.g: NGC 55 (Otte \& Dettmar 1999). It cannot be said that [NII]/H$\alpha$ is constant here  because this ratio is a factor of $2$ larger in the southern H\,{\sc ii} regions than in the rest of the locations, but the differences are not very large, except at the southern part of the galaxy. Actually, more than 60\% of the data range between $0.1-0.2$. Another prediction from the photoionization models is the very large values of this ratio, of 0.3 to 0.6. Such values were not detected anywhere at the DIG of IC 10. A reason might be that these models are optimized for very low ionization values, which is not the situation here. Finally, this ratio can be useful in the determination of the electronic temperature at DIG locations (Miller \& Veilleux 2003). Data in Table 3 have been used to determine the $T_e$ at the 
four cardinal points of the galaxy. Values of 5600~K, 5000~K, 4700~K and 5000~K are found in the south, west, north and east, respectively. The main caveat is the large excitation values of the DIG, which might invalidate the use of the expression (Miller \& Veilleux 2003) $${[NII] \over H\alpha} = 12.2~T_4^{0.426}~e^{-2.18/T_4}$$ where $T_4$ is the electronic temperature in units of $10^4$ K. In order to use this expression it is assumed that there is no N$^{++}$ in the DIG. As the ionization potential of the N$^+$ is close to that of O$^+$, a large excitation might indicate a large fraction of N$^{++}$. Therefore, the former equation might give wrong (lower) values of the T$_e$ than the real ones. 

From the results presented in Table 3 and Figures ~\ref{halfamap}, ~\ref{oiiimap} and ~\ref{choques} that can be concluded there are not large differences in both the excitation and the [NII]/H$\alpha$ ratio throughout the galaxy. Moreover, the value of the [NII]/H$\alpha$ for H\,{\sc ii} regions is large as  compared with other irregular galaxies. This, along with the low values of the [SII]/H$\alpha$ at DIG locations as compared to spiral galaxies, are the characteristics which might be explained by any model. 

\section{Which is the ionization source of the DIG in IC 10?}

It is still under debate if photonization only can explain the line intensity 
ratios observed in the Diffuse Gas in external galaxies . While both [SII]/H$\alpha$ and [NII]/H$\alpha$ ratios can be roughly fitted with models using CLOUDY, the excitation is much more difficult (e.g. Castellanos et al. 04). From the studies carried out so far in both spirals (e.g. Otter \& Dettmar 99) and irregulars (Martin 1997; Hunter \& Gallagher 1992) it is concluded that  the main source of ionization is photoionization from H\,{\sc ii} regions (Greenawalt et al. 1997; Otte \& Dettmar 1999) but shocks can contribute up to 50\% when intense starbursts or a large number of supernova explosions are taking place in the galaxy (Martin 1997). In the present study we come to the conclusion that winds from OB stars do not impel visible shocks in the ISM of IC 10.  On the other hand, from Table 3 it can be concluded that both [NII]/H$\alpha$ and [SII]/H$\alpha$ show lower values in IC 10 than in other spirals but a much larger excitation.

\begin{figure}
\centering
\includegraphics[width=7cm]{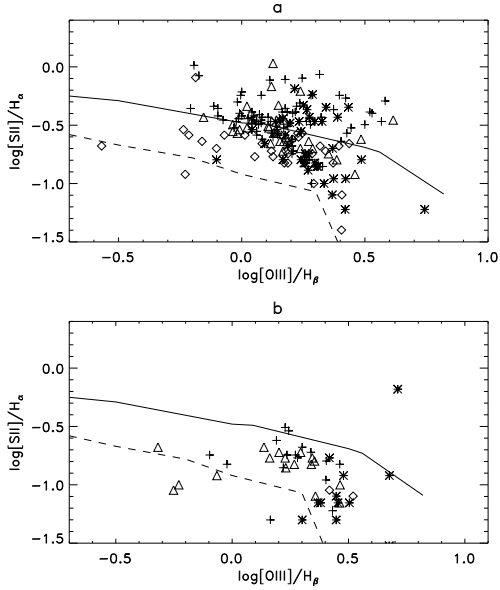}
\caption{  This figure shows the log [SII]/H$\alpha$ vs. log [OIII]$\lambda$5007/H$\beta$.  Different symbols correspond to different cardinal points: crosses (south), stars (west), diamonds (north) and triangles (east). The filled line corresponds to an ionizing temperature of $50,000$ K while a value of $35,000$ K corresponds to the dashed line. All the H\,{\sc ii} regions but three range between these two temperatures but half of the DIG need T$_{ion}$ larger than $50,000$ K.}
         \label{temperatura}
   \end{figure}

In order to explain the high excitation observed in IC 10 with standard photoionization models a $\log~ q$ between $-4$ and $-3$ is needed, with $q$ the ratio between the density of ionizing photons and the electron density,  while values of $-2$ for [SII]/H$\alpha$ and of $0$ for the [NII]/H$\alpha$ ratio are required (Hoppes \& Walterbos 2003). Moreover, the only temperature that can reproduce the excitation is $50,000$ K or larger, while $30,000$ K is enough to explain the other two ratios. This is also seen in Figure ~\ref{temperatura} which shows the log~[SII]/H$\alpha$ vs. log~[OIII]/H$\beta$ diagram for both DIG (a) and H\,{\sc ii} locations (b). Superimposed on the data-points are tracks of ionization temperature from Figure 3 in Martin (1997). This plot is very interesting because the tracks indicate the temperature needed to ionize the medium only by photoionization. According to Sivan, Stasi\`nska \& Lequeux (1986), who used also photoionization models only, an ionization temperature for the DIG of 
35,000 K was obtained for a sample of spiral galaxies as well as different places in the Milky  Way. The situation in IC 10 is very different: in order to ionize half 
of the DIG locations, temperatures larger than 50,000 K are required. Therefore, ionizing stars might be hotter than spectral types O5, which seems quite unreasonable considering only main sequence stars. Moreover, more than half of those locations are at the western side of the galaxy. On the contrary, H\,{\sc ii} locations show a different behaviour. Effective 
temperatures between 35,000 K and 50,000 K, which are typical of H\,{\sc ii} 
regions, can account for the ionization degree. Therefore, it can be concluded that photoionization itself is not enough to obtain the temperatures required for the DIG.

\subsection{Looking for shocks}

\begin{figure}
   \centering
  \includegraphics[width=7cm]{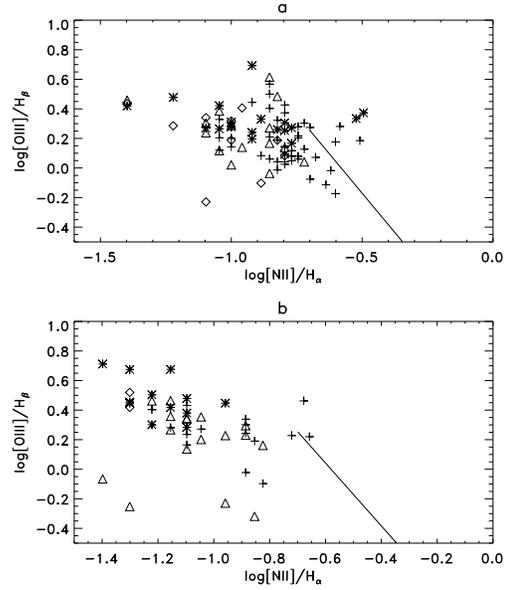}
      \caption{  This figure shows the log [OIII]$\lambda$5007/H$\beta$ vs. log [NII]$\lambda$6583/H$\alpha$ relation. Symbols as in Figure 6. The line separates the plot in two zones: the one to the left where shocks are not important and the zone to the right where shocks play a role. The velocity of the shocks increases from bottom to top of the line.}
         \label{oiiinii}
   \end{figure}

\begin{figure}
  \centering
   \includegraphics[width=7cm]{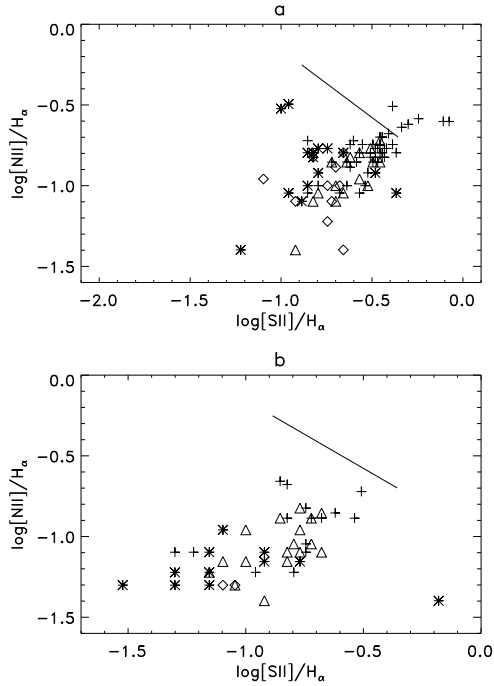}
      \caption{  This figure shows the log [NII]/H$\alpha$ 
vs. log [SII]/H$\alpha$ correlation. Symbols as in Figure 6. The part from the line to the right-up of the diagram corresponds to the shocked zone. Very few of the DIG locations lay inside this region.}
         \label{niisii}
   \end{figure}

\begin{figure}
   \centering
   \includegraphics[width=7cm]{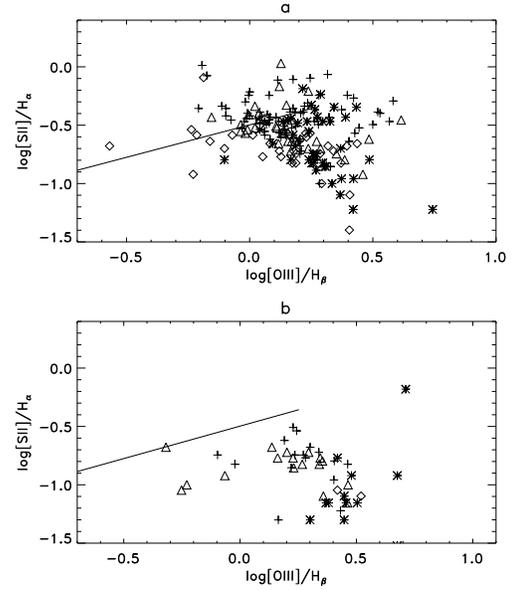}
      \caption{  This figure shows the log~[SII]/H$\alpha$ vs. log~[OIII]$\lambda$5007/H$\beta$. Symbols as in Figure 6. The line divides the  plotted into a shocked part (left-up) and a non-shocked part (right-down).    }
         \label{models}
   \end{figure}

The second most common ionization source for DIG in spiral galaxies is shocks. In order to differenciate between shocked objects (LINERs, SNRs) and H\,{\sc ii} regions, diagnostic diagrams  have been widely used (e.g. Baldwin et al. 1981). In this investigation only those involving the lines detected in all the spectra ([OIII]/H$\beta$, [NII]/H$\alpha$ and [SII]/H$\alpha$) will be used. Together with the data points, the tracks of the models from Dopita \& Sutherland (1995; hereafter DS95) are also plotted in all the diagrams. The lines in Figures ~\ref{oiiinii}, ~\ref{niisii} and ~\ref{models} correspond to the envelope of the models without precursors, a magnetic parameter of $B~n^{-1/2} = 2~ \mu G$ cm$^{-3/2}$ and a shock velocity increasing from 150 km~ s$^{-1}$ to 500 km ~s$^{-1}$ at the left-top. A problem is that these models depend on the 
metallicity and the values in Table 1 in DS95 correspond 
to the metallicity of the Milky Way. IC 10 has a lower metal content and 
when possible, the tracks have been corrected for it.

A well known diagnostic diagram is the log~[NII]/H$\alpha$ vs. log~[OIII]/H$\beta$ which is shown in Figure ~\ref{oiiinii} for DIG (a) and 
H\,{\sc ii} (b) locations. The DS95 models, corrected for the metallicity of IC 10, are also shown. An  
important feature is clear: very few DIG locations lay on the shocked 
part of the diagram (to the right of the line), as all of them are from the south and west sides of the galaxy. Because of the strong dependence of the 
[OIII]/H$\beta$ on the T$_{ion}$  but not on the [NII]/H$\alpha$ ratio, Sivan et al. (1986) concluded that the observational points 
should be aligned on a vertical strip in this plot, when no variations in the nitrogen
content is considered. The vertical column is not clear in this figure, either 
for DIG or H\,{\sc ii} locations. This might indicate variations in the nitrogen or in the ionization temperature at different locations. Nitrogen seems to be homogeneous throughout IC 10 at large scales, but a large variation in T$_{ion}$ (more than $20,000$ K) is needed in other to explain the dispersion in Figure 6a. H\,{\sc ii} regions do not present the vertical strip either (Figure 7b) but a weak anticorrelation. This is more striking because for the H\,{\sc ii} the variations in the T$_{ion}$ are smaller (See Fig. 6b). But, at small scales and inside H\,{\sc ii} regions, nitrogen shows important variations (see paper II). Therefore, a similar result is found, but for different reasons.

Figure ~\ref{niisii} shows the log [NII]/H$\alpha$ vs. log [SII]/H$\alpha$ with the data points and the track of the DS models corrected for the metallicity of IC 10 and with the larger 
magnetic parameter, $2 <$ B~n$^{1/2} < 4 ~\mu$G~cm$^{3/2}$. As in the previous 
diagram, few of the southern DIG locations are in the shocked region, but for all the H\,{\sc ii} and most of the DIG points shocks are not needed. The ratio [NII]/[SII] has been found constant in the halo and disk of spiral galaxies (Haffner et al. 1999; Rand 1998), which cannot be explained by photoionization models (Hoopes \& Walterbos 2003). In Figure ~\ref{niisii}, a trend is clear between these two 
parameters, with  Spearman's regression coeficients of $0.7$ ($\pm~ 0.003$)  and $0.5$ ($\pm ~6 \times 10^{-9}$) for DIG and H\,{\sc ii} locations respectively. This indicates a constancy of the ratio for DIG locations while photoionization is playing a part in the ionization of the H\,{\sc ii} regions, as expected.

Finally, Fig ~\ref{models} shows the log~[SII]/H$\alpha$ vs.~log~[OIII]/H$\beta$ diagram along with the track from the DS95 models at Milky Way metallicity. Although there is no information on how much this line should be shifted for lower metallicities, considering the values for the other two plots it might be less than $0.15$ in log.  That will give us that about $40$ DIG locations lay in the shocked region (now at the left of the line). Half of them corresponds to the southern part of the galaxy. On the contrary, only four H\,{\sc ii} locations lay inside the shocked region.

\begin{figure}
   \centering
   \includegraphics[width=7cm]{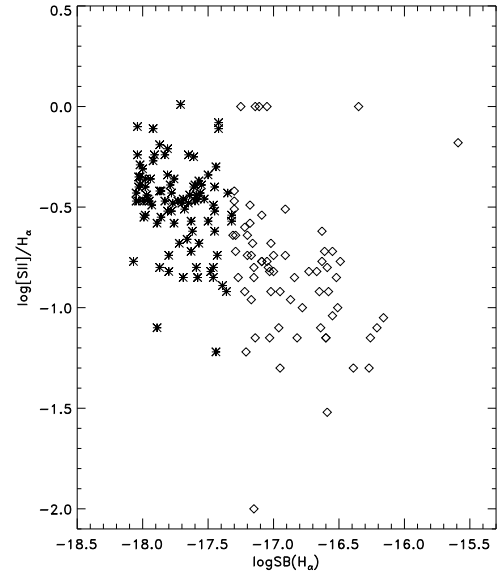}
      \caption{  This figure shows the [SII]/H$\alpha$ vs. log SB(H$\alpha$) for DIG (stars) and H\,{\sc ii} locations (diamonds).  A strong anticorrelation is clear but it is not completely confirmed by the statistics. The vertical cut at log SB(H$\alpha$) =-18.1~mag arcsec$^{-2}$ is due to the reduction problems (See text for details).}
         \label{hasii}
   \end{figure}

\begin{figure}
   \centering
  \includegraphics[width=7cm]{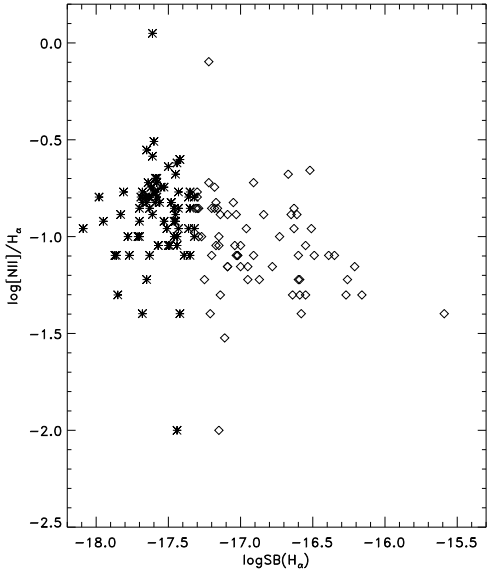}
      \caption{  This figure shows the log~[NII]$\lambda$6583/H${\alpha}$ 
vs. log~SB(H${\alpha})$. Symbols as in Figure 10. An anticorrelation can be inferred from the figure which is not confirmed by the statistics (r$_s$ = $0.7 \pm 0.02$, See text for an explanation). This anticorrelation is forced by the H\,{\sc ii} locations while DIG shows a large dispersion.  }
         \label{hanii}
   \end{figure}

In addition to the diagnostic diagrams, other diagrams can be useful for studying the influence of shocks in the ionization of the DIG.
Martin (1997) in her study of the interstellar medium of irregular galaxies found an anticorrelation between the H$\alpha$ surface brightness and the [SII]/H$\alpha$ ratio which she interpreted as due to shocks. Figure  ~\ref{hasii} shows a similar plot for DIG and  H\,{\sc ii} locations in IC 10. The main differences between Fig. ~\ref{hasii} and Fig. 3 in Martin (1997) are in the surface brightness. They are lower in IC 10 by a factor of $100$. A reason for these differences may be that most of the galaxies in Martin's sample are suffering an intense event of star formation while it is not the case here, as previously discussed. A strong anticorrelation between these two parameters is clear from the figure, but the total regression coefficient is $-0.4$, indicating a weak one, being even weaker for DIG locations ($r=-0.15$). The two-side significance of the deviation from zero of the regression coefficient is $0.22$.  Nevertheless, the dispersion is very high. In Figure  ~\ref{hasii} there is a ``plume'' at log~[SII]/H$\alpha$ = $0.0$, mainly of H\,{\sc ii} locations. This might resemble the bifurcation seen in Martin's figure at log~SB(H$\alpha$) = -14.2, which can account for the differences in metallicity among the galaxies in her sample. The main caveat is that apparently there are not important differences in the metallicity throughout the galaxy (see paper II). Some authors say that log~[NII]/H${\alpha}$ vs. log~SB(H${\alpha}$) should show a similar behaviour to log~[SII]/H${\alpha}$ vs. log~SB(H${\alpha}$) because both are enhanced by the same mechanisms (e.g. Slavin 2003). Figure ~\ref{hanii} shows that there is no correlation between log~[NII]/H$\alpha$ and log~SB(H${\alpha}$). Actually, the regression coefficient is $0.02~$ (with a deviation of $\pm~0.02)$ for the total sample. This lack of correlation is not due to the uncertainties in this ratio. One explanation could be related with the large WR population. As these stars  among other chemical elements such carbon and helium (Hamann 1995) release nitrogen to the ISM, the content of this element inside IC 10 might be large, even if the released amount of nitrogen is not very large, and, more important here, is inhomogeneously distributed throughout the galaxy, with larger values in the neighbourhood of these stars. Favouring this explanation is the large [NII]/H$\alpha$ value and the large 12+log(N/H) as well as the possible inhomogeneities in nitrogen at low scales (see paper II). 

From Figures ~\ref{oiiinii}, ~\ref{niisii}, ~\ref{models}, ~\ref{hasii} and ~\ref{hanii} it can be concluded that shocks as a secondary ionization source are only needed for a few points located at the southern side of IC 10, but at the same time, photoionization is not the only source for the rest of the locations. 

\subsection{The leaking model}

\begin{figure}
   \centering
   \includegraphics[width=7cm]{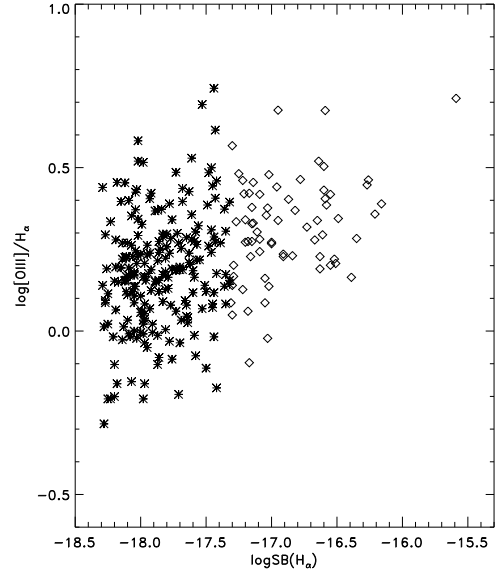}
      \caption{    This figure shows the log [OIII]$\lambda$5007/H$\beta$ 
vs. log SB(H$\alpha$) relationship.  Symbols as in Figure 10. A low envelope is clear from this figure while the upper one is almost flat. }
         \label{oiiiha}
   \end{figure}

In order to know which is the real ionization source for the DIG, the excitation could be very useful because [OIII]/H$\beta$ is very sensitive to the chemical composition (Searle 1971, but See McGaugh 1994 for a different interpretation) but also to the ionization parameter as an indicator of the ionizing photon density to the electronic density.  Firstly, we can study the relation between the excitation and the H$\alpha$ flux. If the latter is related with the flux of the ionizing photons, a correlation between these parameters can be expected. From Figures ~\ref{halfamap} and ~\ref{oiiimap} a preliminary answer can be inferred and is actually confirmed in Figure ~\ref{oiiiha}. This Figure shows the relationship between [OIII]/H$\beta$ and the SB in H$\alpha$ for the DIG and the H\,{\sc ii} locations in IC 10. A  regression coefficient of 0.4 is obtained for the total sample with a deviation from zero of $10^{-11}$. Such a weak correlation can be interpreted in several ways: the chemical oxygen  abundances are not very homogeneous (see paper II for a discussion) but the explanation based on changes in the ionization parameter is more likely. These changes can be in the electronic density or in the ionizing photon density. The data presented here give no information about the electronic density but changes in the ionization parameter  can indicate an extra source of ionization from the OB stars, as previously mentioned. In this situation the WR stars population in IC 10 must be taken into account.  This is in agreement with the results from Garlaza, Walterbos \& Braun (1999), who interpreted the lack of correlation between the H$\alpha$ flux and the excitation in M31, as well as the large value of the excitation, as a clear indication of a secondary source of ionization for the ISM
 
Moreover, as shown in Table 3, the excitation is very large in the DIG locations which means a very hard spectrum. There are several reasons for it. A spectrum emerging from H\,{\sc ii} regions could be apparently harder due to photon leaking (Hoopes \& Walterbos 2003). The larger
the leaking, the harder the spectrum seems to be. A harder spectrum means larger values in the excitation, with cooler stellar temperatures. There is some improvement when this leakage of photons is considered. Both the temperature and the ionization parameter for the excitation and the [SII]/H$\alpha$ ratio are similar, $40,000$K and $-3$ respectively, with a fractional leakage of $40$ \%. But, the [NII]/H$\alpha$ ratio does not fit any of these models. The only solution could be a low ionization parameter, that would work against the large excitation. Therefore, a combination of leaking H\,{\sc ii} regions and a high ionization parameters can account for the low values in [SII]/H$\alpha$ and the high excitation, but none of the values of $q$ or leakage can explain the [NII]/H$\alpha$. 

It can be argued that the large excitation could be due to other reasons. Low metallicity regions tend to produce higher temperatures for the same hardness of the spectra (Cox, private communication) which would explain the large values of the ionization temperature required in Figure ~\ref{temperatura} and there is no need for very hot stars. But there are galaxies with lower metallicity and lower excitation than IC 10 (e.g., DDO 50, Hidalgo-G\'amez et al. in preparation). Another argument could be that OB stars themselves can excite the ISM as they have winds extendeding over a long time. Unfortunatelly, it is very difficult to disentangle the contribution of the winds from WR stars and from  OB stars since both kinds of stars are generally located in the same regions: WR stars release stronger winds than OB stars, albeit over shorter times.

Finally, another explanation is that the spectrum is really a hard one because of the existence of many hot stars. In IC 10 the perfect candidates are the WR stars. They have large $T_{ion}$ and due to the large population they will provide a large ionization parameter. It is presumed that there are at least $100$ of such stars in 
this galaxy (Massey \& Holmes 2002). As they are hotter than OB stars their ionization spheres are also larger.  Each Wolf-Rayet star produces a Str\"omgren sphere larger than $125$ pc and a flux of $7\times 10^{39}$~ erg~ s$^{-1}$ (Hidalgo-G\'amez 2005). From Figure 13 in Wilcots \& Miller (1998), $10$ out of the $20$ WR stars detected in the galaxy are located inside the region under study which has an area of $147,900$ pc$^2$. The area covered by the Str\"ongrem spheres of these $10$ WR stars is about $122,719$ pc$^2$. Therefore, it is  enough to produce the high and homogenous excitation observed in this part of the galaxy. On the other hand, due to their short life time ($\approx 10^5$ yr), each of these stars can produce at most $10\%$ of the ionizing photons needed for the DIG ionization level (Hidalgo-G\'amez 2005), but again with $10$ stars in such small area the number of ionizing photons is sufficient to explain the excitation. 

Therefore, the best explanation for the hardness of the spectra, the relatively large content of nitrogen (as compared with other dwarf irregular galaxies) and the lack of shocks is the ionization 
by the leaking from the H\,{\sc ii} regions in combination with the ionization from the large population of WR 
stars.

\section{Conclusions}

The ionized diffuse medium in IC 10 shows very different features from those observed in spiral galaxies. Especially the high excitation as well as the low values of [SII]/H$\alpha$ are very noticeable and might be related.  The other interesting ratio, [NII]/H$\alpha$, is not only different from all the values obtained in spiral galaxies but it cannot be fitted with any photoionization model. A leaking of $40 \%$ of the photons created by massive stars and a large population of WR stars can explain the intensities of the ratios observed for most of the locations. For those needing an extra source, mainly some locations towards the south of the galaxy, shocks induced by winds from the very massive, hot WR stars can be proposed. However, there are a few problems to be solved. 

The most important is that though the velocities of the WR winds are larger than those from the OB associations, the [SII]/H$\alpha$ ratio in not enhanced sufficiently to be considered as shocked. Actually, less than a dozen of locations 
shows values of this ratio larger than 0.7, which is the value obtained from the  Sutherland \& Dopita (1993) models for velocities of $500$ km s$^{-1}$ and very low densities. Moreover, such winds  were not detected for OB winds. There are several facts favouring the wind hypothesis: the shock waves related with WR stars 
are confined to a small region around the stars, (20 pc, Esteban, private communication) which agrees with the fact that not all the locations in the south need an extra source of ionization by shocks. The main caveat is that the majority of the WR stars detected is located in the eastern part of IC 10 (see Fig. 1 in Massey et al. 1992).  

Although these two problems, need more accurate data and better resolution to be solved, the best explanation for the spectral characteristic of the DIG is photon leaking from  the H\,{\sc ii} regions in combination with the ionization from the large population of hot WR 
stars. 

\begin{acknowledgements}

A.M.H-G.~ is indebted to M. Richer and A. Bullejos who acquired the data. 
She wants to thank to M. Richer for many lively discussions as well as many 
comments and suggestions which really improve this investigation. Also, L. 
Georgiev is thanked for his help and many comments on this work and
C.~Morisset for his help in handling the figures. She also 
thanks to Uppsala Astronomiska observatoriet for making the program PHYZ 
available, and N. Bergvall for the metallicity software. M. Castellanos, 
C. Esteban, M. Peimbert, M. Rosado and P. V\'{\i}lchez are thanked for 
fruitful discussions. 
This work was partly supported by CONACyT project 2002-C40366. 

\end{acknowledgements}

\end{document}